\documentclass[letterpaper, 10 pt, conference]{ieeeconf}  

\IEEEoverridecommandlockouts                              

\usepackage{graphics} 
\usepackage{epsfig} 
\usepackage{mathptmx} 
\usepackage{times} 
\usepackage{amsmath} 
\usepackage{amssymb}  
\let\labelindent\relax
\usepackage{enumitem}
\usepackage{bm}
\usepackage{xcolor}  
\usepackage{subfigure}
\usepackage{graphicx}
\usepackage{colortbl}
\usepackage{censor}
\usepackage{todonotes}
\usepackage{lipsum}
\usepackage{cite}
\usepackage{url}
\usepackage{multirow}
\usepackage{xcolor}
\usepackage{optidef}
\usepackage{arydshln}
\usepackage[hidelinks]{hyperref}
\usepackage{algorithm}
\usepackage{hyperref}
\usepackage{siunitx}
\usepackage{booktabs}
\usepackage[noEnd=true]{algpseudocodex} 
\DeclareMathOperator*{\minimize}{minimize}

\usepackage[font=small,skip=3pt]{caption}

\title{\LARGE \bf
Toward a Decision Support System for Energy-Efficient Ferry Operation on Lake Constance based on Optimal Control}

\author{Hannes Homburger$^{1}$, Bastian Jäckl$^{2}$, Stefan Wirtensohn$^{1}$, Christian Stopp$^{1}$,\\ Maximilian T. Fischer$^{2}$, Moritz Diehl$^{3}$, Daniel A. Keim$^{2}$, and Johannes Reuter$^{1}$
\thanks{$^{1}$ Institute of System Dynamics, HTWG Konstanz, 78462 Konstanz, Germany,
         {\tt\small hhomburg@htwg-konstanz.de}}
    \thanks{$^{2}$ Department of Computer and Information Science, University of Konstanz, 78464, Konstanz, Germany}%
    \thanks{$^{3}$ Department of Microsystems Engineering (IMTEK) and Department of Mathematics, University of Freiburg, 79110 Freiburg, Germany}
}
\begin{document}
\maketitle
\thispagestyle{empty}
\pagestyle{empty}

\begin{abstract}
The maritime sector is undergoing
a disruptive technological change driven by three main factors: autonomy, decarbonization, and digital transformation.
Addressing these factors necessitates a reassessment of inland vessel operations. This paper presents the design and development of a decision support system for ferry operations based on a shrinking-horizon optimal control framework. The problem formulation incorporates a mathematical model of the ferry's dynamics and environmental disturbances, specifically water currents and wind, which can significantly influence the dynamics. Real-world data and illustrative scenarios demonstrate the potential of the proposed system to effectively support ferry crews by providing real-time guidance. This enables enhanced operational efficiency while maintaining predefined maneuver durations. The findings suggest that optimal control applications hold substantial promise for advancing future ferry operations on inland waters. A video of the real-world ferry \textit{MS Insel Mainau} operating on Lake Constance is available at: {\fontfamily{qcr}\selectfont https://youtu.be/i1MjCdbEQyE }
\end{abstract}

\section{Introduction}
Recent developments in the field of numerical optimal control, combined with the steadily increasing embedded computing power, enable innovation in autonomy, decarbonization, and digital transformation of inland vessel operation. Pioneered by Gilles and coworkers in the 1990s \cite{Wahl.1998}, this innovation can come in the form of advanced optimal control approaches in inland vessel scenarios. 
Recently, advanced control approaches have been shown to efficiently control vessels autonomously in real-world scenarios \cite{Lexau.2023}. However, regulatory constraints, concerns about safety in edge cases, and questions of economic efficiency currently limit large-scale deployment. In passenger transport, stakeholder trust, explainable system behavior, and profitability are essential factors to achieve real-world impact \cite{Lutzhoft.2024}. These factors are particularly relevant for ferries carrying hundreds of passengers, where there remains a strong preference for a human captain on board. Captains often face challenges when dealing with environmental disturbances such as currents and winds exhibiting non-intuitive characteristics. This motivates the implementation of a decision support system to reduce operator workload and enhance economic efficiency. In this paper, we present the design of a decision support system for the commercial electric ferry \textit{MS Insel Mainau} capable of transporting up to 300 passengers on Lake Constance.

\subsection{Contributions}
\begin{figure*}[t!]
 \vspace*{0.3cm}
	\centering
	\includegraphics[width=0.85\textwidth]{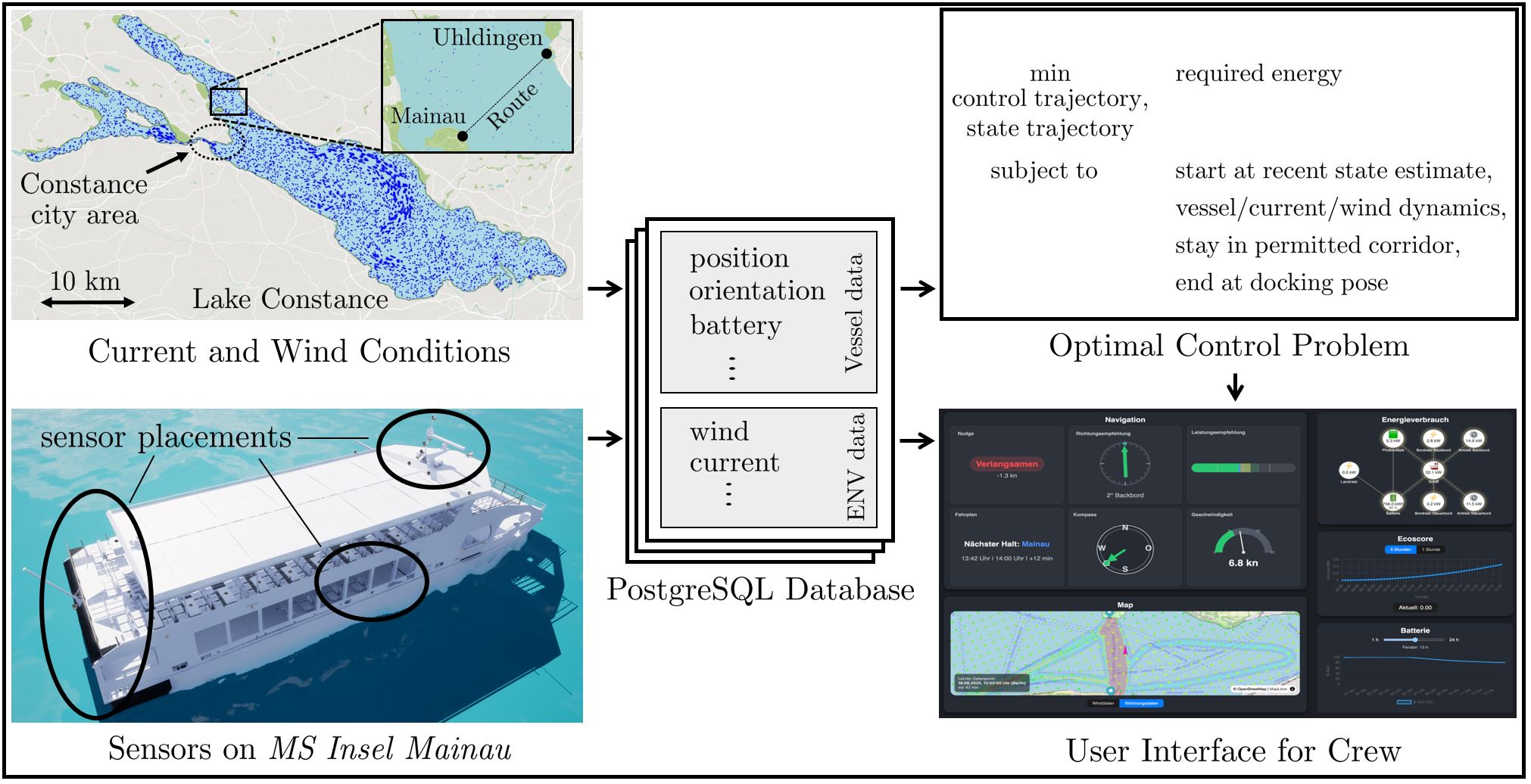}
	\caption{Schematic architecture of the decision support system for the real-world ferry \textit{MS Insel Mainau}, illustrating the sequential flow from data collection (left), through data storage and management (center), to optimization and visualization modules (right).}
	\label{fig_Schematic_Visu}
 \vspace*{-0.3cm}
\end{figure*}

The following main contributions toward energy-efficient operation are presented in this paper:

\begin{itemize}
    \item \textbf{Shrinking-Horizon Optimal Control for Real-World Ferry Operation:} A novel decision support system is proposed based on a shrinking-horizon optimal control problem formulation, which incorporates environmental influences such as water currents and wind. The system employs an efficient direct optimal control approach suitable for providing real-time solutions.

    \item \textbf{Digital Twin of the Ferry and Environmental Disturbances:} A comprehensive mathematical model of the fully-electric ferry \textit{MS Insel Mainau} is developed, including parameter identification and evaluation. Additionally, an optimization-friendly model architecture for current and wind disturbances is selected and validated using data from a PDE-based model.

    \item \textbf{Implementation of the Decision Support System:} The system plans energy-optimal trajectories with respect to the environmental conditions, preserving predefined maneuver durations. This illustrates the potential to support ferry crews with actionable guidance, highlighting the method’s practical feasibility for real-world deployment.
\end{itemize}
 A schematic visualization of the proposed architecture of the decision support system is provided in Fig.~\ref{fig_Schematic_Visu}.

\subsection{Requirements and Related Work} 

\label{sec_req}
To increase the efficiency of vessel operations and reduce emissions, replacing conventional marine fuel with alternatives \cite{Bouman.2017} or reducing operation speed \cite{Schubert.2024} are prevalent measures. Complementarily, numerical optimization is a promising tool for further improving ferry operations \cite{Barreiro.2022}, particularly in tough
environmental conditions \cite{Homburger.2024}. 
The primary objective of the decision support system design is to provide the crew with valuable information and insights. Besides an intuitive visualization of the most important processing data, suggestions for the optimal route and velocity profile are essential to the crew. These suggestions should consider the dynamics of the ferry and the following \textbf{\mbox{requirements}}: 
\begin{itemize}
    \item Stay in the permitted corridor.
    \item End at the docking pose at the scheduled time.
    \item Minimize required energy.
\end{itemize}
 The successful fulfillment of these requirements is essential to ensure acceptance by crew, thereby enabling meaningful real-world impact~\cite{Lutzhoft.2024}.
 
A \mbox{\textbf{selection of related work}} is given by the following research projects considering (semi-)autonomous real-world ferries. Important examples are the \texttt{Zeabuz} project in Stockholm \cite{Hjelmeland.2022}, the autonomous ferry prototypes \texttt{milliAmpere} \cite{Hinostroza.2025} and \texttt{milliAmpere2} \cite{Alsos.2024} from NTNU in Norway, the \texttt{GreenHopper} \cite{Enevoldsen.2022} from DTU in Denmark, the German projects \texttt{CAPTN} with the prototype \texttt{MS WaveLab} \cite{AlFalouji.2023} in Kiel, and the \texttt{Akoon} project with the river ferry \texttt{Horst} in Ingelheim \cite{Koschorrek.2022}. \textbf{Weather-routing} tools for vessel operation range from tree-search methods \cite{Bentin.2016} over dynamic programming \cite{Koschorrek.2022} to A$^\star$-based algorithms \cite{Grifoll.2022}. Empirical evidence for ocean freight indicates that weather routing can yield considerable fuel savings \cite{Bentin.2016}. A comprehensive survey of weather-routing systems for vessels is provided in \cite{Zis.2020}. In contrast to previous studies, we propose a system designed for inland waterway operation characterized by limited route length and fixed duration. This setting enables a novel, environment-aware direct optimal control formulation, which includes a dynamic ferry model and potentially provides more interpretable and efficient suggestions.

\subsection{Outline} In Section~\ref{sec_overview}, an overview of the system is given and the shrinking-horizon optimal control problem is stated, which is the basis for the decision support system. This is followed by a presentation of the ferry's dynamics, the power model, and the limited operation corridor. To simplify the vessel dynamics, observations from real-world operation of the ferry \textit{MS Insel Mainau} are collected and exploited. Section~\ref{sec_exp} presents the numerical results based on real-world data. The validation contains a descriptive data analysis of the environmental conditions, an empirical parameter identification of the ferry model, and two illustrative case study scenarios.  Finally, Section~\ref{sec_con} concludes the paper and discusses ideas for future work.

\section{Architecture of the Support System}\label{sec_overview}
To meet the requirements defined in the previous section, the decision support system is implemented in the modules data collection, data storage and management, optimization, and visualization.
A schematic illustration of the architecture is given in Fig.~\ref{fig_Schematic_Visu}.
Data collection comprises 92 data streams on the \textit{MS Insel Mainau}, recorded at 5-second intervals in a PostgreSQL database. The processing data of the ferry is measured using mounted sensors, most prominent:
\begin{itemize}
\item Pose, including GPS-position and angular orientation.
\item Propellers, including speed, orientation, and power.
\item Energy flow, including state of charge of the batteries and solar power supply. 
\end{itemize}
Additionally, hourly current and wind data on Lake Constance are tracked. Those are calculated by the numerical solution of a system of partial differential equations (PDEs) presented in previous work \cite{Lang.2010}. Selected data is visualized in a user interface (UI) for the captain and also fed into an optimal control problem, which predicts energy-efficient trajectories under the given constraints. The resulting trajectories and self-explanatory nudges are then visualized. In the next part, we detail the optimal control problem formulation.

\subsection*{Optimal Control Problem (OCP) Formulation}\label{sec_prob}
\begin{figure}
 \vspace*{0.2cm}
	\centering
	\includegraphics[width=0.49 \textwidth]{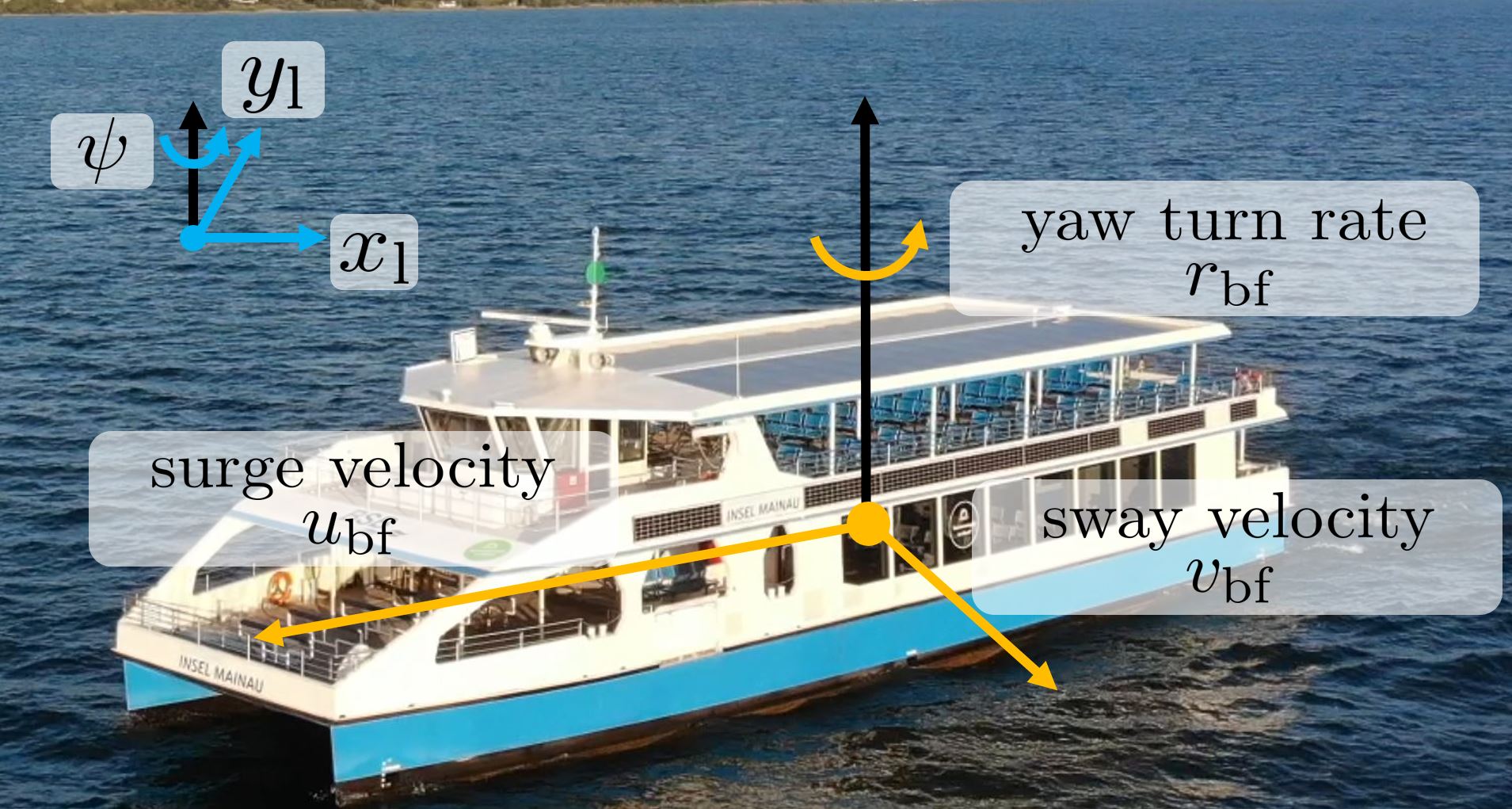}
	\caption{ Photography of the ferry \textit{MS~Insel~Mainau} on Lake Constance, including the 3-DOF velocity in the body-fixed frame (gold) and the 3-DOF pose in the ENU frame (blue).}
	\label{fig_frames}
    \vspace{-0.45cm}
\end{figure}

To compute optimal input and state trajectories for ferry maneuvers, the defined requirements are embedded in the shrinking-horizon OCP: 
\begin{subequations}\label{eq_ocp}
\begin{align}
    \minimize_{u(\cdot),\:x(\cdot)} \hspace{0.4cm}\int_t^T P(u(\tau)&)+\lvert| Sx(\tau) \rvert|_Q^2 + \lvert| u(\tau) \rvert|_R^2 \;\mathrm{d}\tau \label{eq_ocp_objective}\\
    \mathrm{subject\; to}\hspace{0.8cm} x(t)-\hat x_t&=0,\label{eq_ocp_initial}\\
    x(T)-x_\mathrm{Dock}&=0,\label{eq_ocp_terminal}\\
    \dot x(\tau)-f(x(\tau),u(\tau),\xi)&=0, \mathrm{\;\;\;for\;} t\leq\tau\leq T,\label{eq_ocp_dynamics}\\
     h(x(\tau),u(\tau))&\leq 0,   \mathrm{\;\;\;for\;} t\leq\tau\leq T.\label{eq_ocp_path}
\end{align}    
\end{subequations}
 The decision variables are the input trajectory $u:[t,T]\rightarrow\mathbb{R}^{n_u}$, the state trajectory $x:[t,T]\rightarrow\mathbb{R}^{n_x}$, where the end time of the maneuver is $T$. The exogenous parameters of the OCP are the recent time $t\leq T$, the recent state estimate $\hat x_t\in\mathbb{R}^{n_x}$, and the vector $\xi\in\mathbb{R}^{n_\xi}$ that parameterizes the current and wind model. Note that during operation, the time $t$ increases and the duration of the prediction horizon shrinks. 
The docking pose is denoted by $x_\mathrm{Dock}\in\mathbb{R}^{n_x}$. The objective of the OCP \eqref{eq_ocp_objective} is given by the required energy, computed by the time integral of the power $P$ corresponding to a maneuver in the current and wind field, and additional quadratic regularization terms on the control inputs and the body-fixed velocities (cf. Fig.~\ref{fig_frames}), which are selected from the state vector with the matrix $S=\left[0
_{3\times 3}, I_{3\times 3}\right]$. The weighting matrices $Q,R\succeq0$ are positive semi-definite. The initial state of the maneuver is fixed to the recent state estimate in \eqref{eq_ocp_initial}. The terminal state is fixed to the docking position \eqref{eq_ocp_terminal}. The dynamical feasibility of the trajectory is ensured by the equality constraints \eqref{eq_ocp_dynamics}. The path constraints \eqref{eq_ocp_path} limit the permitted corridor for operation and the admissible inputs of the system. Note that a suitable choice of $T$ ensures the ferry's arrival at the docking pose at the scheduled time.
Note that the OCP \eqref{eq_ocp} can be referred to as
$\mathcal{P}(t,\hat x_t, \xi)$ that emphasizes the problem's dependency on its exogenous parameters. In the following, we outline the parts of the OCP. 

\subsection{Dynamic model of the ferry}
\begin{figure}[t!]
 \vspace*{0.2cm}
	\centering
	\includegraphics[width=0.39\textwidth]{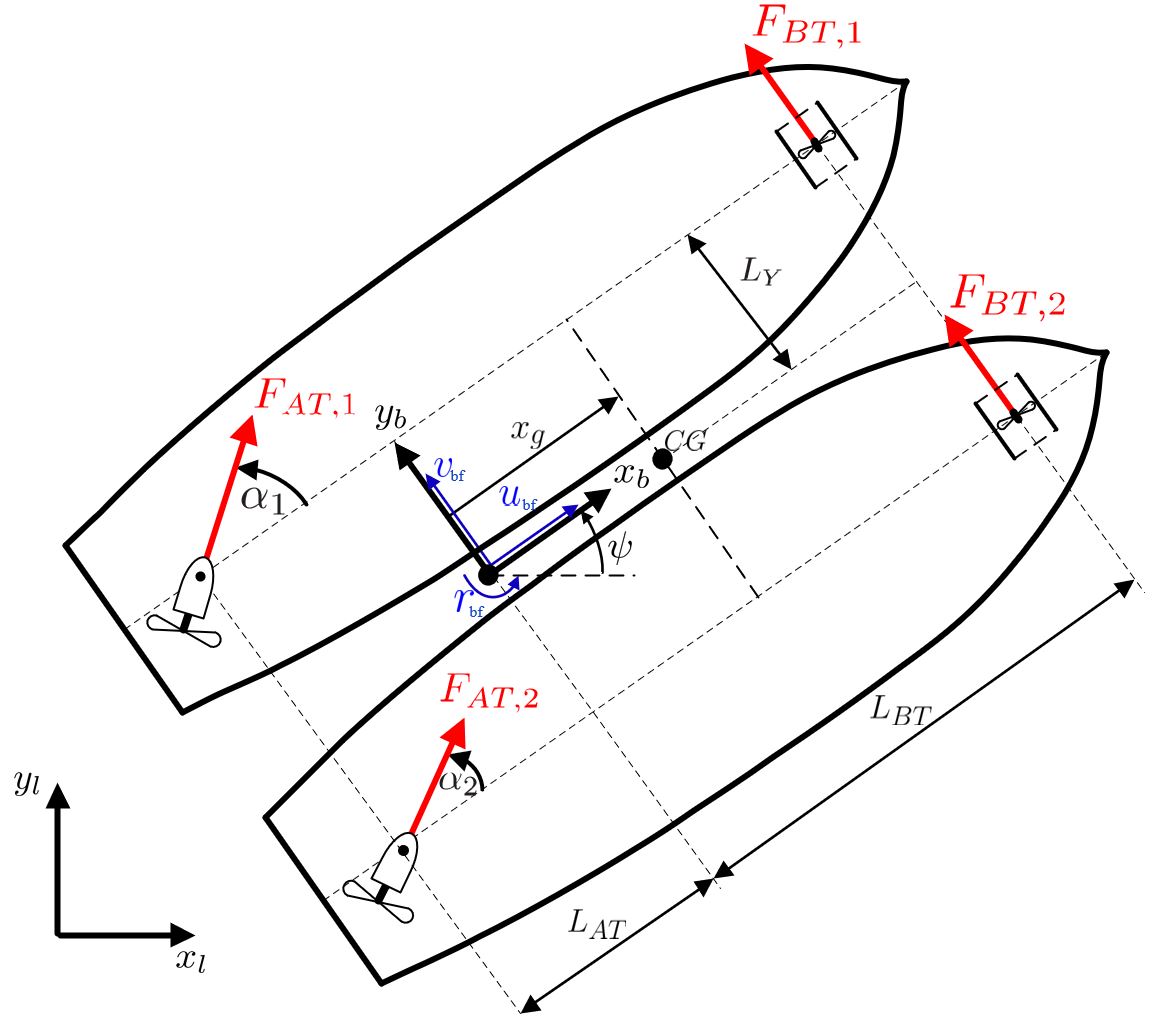}
	\caption{ Schematic drawings of the reduced 3-DOF state of the \textit{MS~Insel~Mainau} including the actuators (red).}
	\label{fig_schematic}
    \vspace{-0.45cm}
\end{figure}

 In inland water scenarios, pitch, roll, and heave motions are small and can be neglected \cite{Fossen.2021}. Therefore, the dynamics of the vessel can be described by a nonlinear three-degree-of-freedom (3-DOF) maneuvering model, as a special case of \textit{Fossen's equation} given by

\begin{subequations}
\label{eq_3DOF_eta_dyn}
\begin{align}
    \dot{\eta}&=J_\psi(\eta)\nu,\\
    M\dot\nu&=\tau_\mathrm{a}(a)+\tau_\mathrm{w}(\nu,\xi,\eta)  -C(\nu)\nu-D(\nu,\xi,\eta).
\end{align}    
\end{subequations}
In contrast to the standard presentation given in \cite{Fossen.2021}, a dependence of the hydrodynamic damping $D$ and the wind disturbance $\tau_\mathrm{w}$ on the parameter vector $\xi$ and the pose $\eta$ is given, which is used to model the influence of \textit{spatial} environmental effects. 
The state of the dynamic model \eqref{eq_3DOF_eta_dyn} is given by the local 2D-position and yaw angle with respect to the $x_\mathrm l$-axis collected in $\eta=[x_\mathrm{l},y_\mathrm{l},\psi]^\top$ in the east-north-up (ENU) frame and $\nu=[u_\mathrm{bf},v_\mathrm{bf},r_\mathrm{bf}]^\top$ that denotes the body-fixed velocity vector over ground. The homogeneous 2D rotation matrix $J_\psi(\eta)$ transforms from the body-fixed to the ENU frame, $\tau_\mathrm{a}$ denotes the forces and torque in the body-fixed frame applied by the actuators, and the positive definite mass matrix is $M\in\mathbb{R}^{3\times 3}$. The Coriolis and centripetal effects are embedded in the matrix  $C:\mathbb{R}^3\rightarrow\mathbb{R}^{3\times 3}$. The control inputs and the 3-DOF state of the \textit{MS Insel Mainau} are visualized in Fig.~\ref{fig_schematic}. \\


\subsection{Model simplification for real-world application}
While a homotopy-based approach to plan energy-optimal docking trajectories in strong currents with a full 3-DOF vessel model is presented in previous work \cite{Homburger.2024}, the computational effort required by this approach is often too high for real-time application. Therefore, the knowledge on the real-world operation of \textit{MS Insel Mainau} is exploited to simplify the dynamics model.
The analysis of real-world operations reveals the following observations:
\begin{enumerate}[label=(\alph*)]
\item The propeller speeds and the orientations of the two azimuth thrusters are roughly identical during operation.
\item The bow thrusters are only used in extreme cases.
\item The angle of attack (AoA) is crucial for the required \mbox{energy,} and should be chosen depending on environmental conditions.
\item The body-fixed velocity profile has a high impact on the required energy.
\item The turn rate of the vessel's yaw angle is very low.
\end{enumerate}
 The system input is $a=[F_\mathrm{AT1},\alpha_\mathrm{1},F_\mathrm{BT1},F_\mathrm{AT2},\alpha_\mathrm{2},F_\mathrm{BT2}]^\top\in\mathbb{R}^{6}$ (cf. Fig.~\ref{fig_schematic}) and represents the states of the actuators. Based on observations (a)-(b), we assume
$F_\mathrm{BT,1}=F_\mathrm{BT,2}=0$, $\alpha_1=\alpha_2=:\alpha$, $F_\mathrm{AT,1}=F_\mathrm{AT,2}=:F_\mathrm{AT}$, resulting in the simplified control allocation given by 
\[X_\mathrm a=2F_\mathrm{AT}\cos(\alpha) \quad\mathrm{and }\quad Y_\mathrm a=2F_\mathrm{AT}\sin(\alpha).\]
Based on an interplay of observations (c)-(e), we assume direct control over the actuator force in surge direction $X_\mathrm a$, the actuator force sway direction $Y_\mathrm a$, and the turn rate of the vessel $\dot r_\mathrm{bf}$, yielding the control input
$ u =[ X_\mathrm a,  Y_\mathrm a,\dot r_\mathrm{bf}]^\top $, and the dynamics are simplified to
\begin{subequations}\label{eq_ode}
\setlength{\arraycolsep}{1.5pt}
\begin{align}
\dot{\eta}&=J_\psi(\eta)\nu,\\
\dot \nu&=\begin{bmatrix}
 m^{-1}\!\bigg(
 \begin{bmatrix} X_\mathrm{a}\\ Y_\mathrm{a}\end{bmatrix}
 +\tilde\tau_\mathrm{w}(\eta,\nu,\xi)
 -\tilde C(\nu)\nu
 -\tilde D(\eta,\nu,\xi)
 \bigg)\\
 \dot r_\mathrm{bf}\label{eq_nu}
\end{bmatrix},
\end{align}
\end{subequations}
where the terms superscripted by $\sim $ contain the first and second elements of the corresponding terms without the superscript.
   The simplified ODE system \eqref{eq_ode} can be rewritten in the standard form $\dot{x}=f(x,u,\xi)$ using the state vector $x=\left[\eta^\top,\nu^\top\right]^\top\in\mathbb{R}^{6}$ and is part of the shrinking-horizon OCP~\eqref{eq_ocp_dynamics}. 
\subsection{Local current and wind model}
As discussed at the beginning of this section, the environmental conditions are collected in the form of the numerical solution of a system of PDEs \cite{Lang.2010}. This numerical solution is given for the entire surface area of Lake Constance in the form of tabular data tuples with entries in the form of
\begin{equation}\label{eq_tab_data}
    [x_\mathrm{l},y_\mathrm{l},v_{x,\mathrm{wind}},v_{y,\mathrm{wind}},v_{x,\mathrm{current}},v_{y,\mathrm{current}}]^\top\in\mathbb{R}^6. 
\end{equation} 
This representation of the environmental conditions is simple but not optimization-friendly in the sense that first- and second-order derivatives are not provided. Further, as depicted in the upper left illustration in Fig.~\ref{fig_Schematic_Visu}, the ferry operates only in a limited area of Lake Constance. Therefore, a local model of the environmental conditions, even providing first- and second-order information, is a key element for efficient optimization.  
For this aim, a spatial second-order model is chosen and specified as
\begin{equation}\label{eq_env_model}
    \left[\begin{array}{c}
         v_x(x_\mathrm{l},y_\mathrm{l})  \\
           v_y(x_\mathrm{l},y_\mathrm{l}) 
    \end{array} \right]= \left[
    \begin{array}{c}
    \tfrac{1}{2}p^\top Q_\mathrm{x}p+L_\mathrm{x}p+\mu_\mathrm{x}\vspace*{0.1cm}  \\
    \tfrac{1}{2}p^\top Q_\mathrm{y}p+L_\mathrm{y}p+\mu_\mathrm{y}
    \end{array} \right],
\end{equation}
where $p:=[x_\mathrm{l},y_\mathrm{l}]^\top$ 
is the 2D-position on the water and $Q_\mathrm{x},Q_\mathrm{y}\in\mathbb{R}^{2 \times 2}$, $L_\mathrm{x},L_\mathrm{y}\in\mathbb{R}^{1\times 2}$, and $\mu_\mathrm{x},\mu_\mathrm{y}\in\mathbb{R}$ contain the parameters. Since $Q_\mathrm{x}$ and $Q_\mathrm{y}$ can be assumed symmetric without loss of generality, the quadratic models of the 2D current field and 2D wind field require 12 independent parameters each. Consequently, the total number of parameters is $n_\xi = 24$.
Note that the introduced spatial quadratic models are optimization-friendly and have low complexity. 
\subsubsection*{Environmental effects on the vessel dynamics}
The hydrodynamic damping effects are modeled in $\tilde D$ and the force vector by the relative motion in a wind field is $\tilde \tau_\mathrm{w}$, both dependent on the parameter vector $\xi$. 
In detail, the relative velocity to the surrounding water contain the components $u_{r,\mathrm{water}}$, $v_{\mathrm{r,water}}$ with AoA $\gamma_{\mathrm{r,water}}$ and the relative velocity to the surrounding air is given by the components $u_{r,\mathrm{wind}}$, $v_{\mathrm{r,wind}}$ with AoA $\gamma_{\mathrm{r,wind}}$. To model the hydrodynamic damping effects, we choose a second-order modulus model given in \cite[Sec.~4]{Homburger.2025}. 
To model the wind effects, a standard wind model \cite[Sec.~10]{Fossen.2021} given by
\begin{equation*}
    \tilde \tau_\mathrm{w}=\frac{1}{2}\rho \left(u_\mathrm{r,wind}^2+v_\mathrm{r,wind}^2\right)\left[\begin{aligned} -c_x&\:\cos(\gamma_\mathrm{r,wind})\:A_\mathrm{Fw}\\ c_y&\:\sin (\gamma_\mathrm{r,wind})\:A_\mathrm{Lw} \end{aligned}\right]
\end{equation*}
is employed, where $\rho$ is the air density, $c_x,c_y\in[0,1]$ are drag coefficients, and $A_\mathrm{Fw},A_\mathrm{Lw}>0$ are projected areas. Note that no torque is considered in this model, as we assume direct control over the vessel's turn rate in Eq.~\eqref{eq_nu}.  \\
\subsection{Power model}
The power required by the actuators of the ferry is 
\begin{equation*}
    P(u) =2c_\mathrm p\left| n_\mathrm{AT}^3\right|,
\end{equation*}
where $c_\mathrm p>0$ is a coefficient and $n_\mathrm{AT}:=n_\mathrm{AT,1}=n_\mathrm{AT,2}$ are the propeller speeds of the azimuth thrusters (AT) required to provide the forces $X_\mathrm{a}$ and $Y_\mathrm{a}$, respectively.
Using the simplified control allocation, we obtain
$4F_\mathrm{AT}^2=X_\mathrm{a}^2+Y_\mathrm{a}^2$.
By ignoring efficiency changes given by slip due to the ferry's moderate velocity, the relation $F_\mathrm{AT}\propto n_\mathrm{AT}^2$ holds and  \[ P(u)= 2 \check c_\mathrm p  \left( \frac{1}{4}(X_\mathrm{a}^2+Y_\mathrm{a}^2)\right)^{\tfrac{3}{4}} \]
is obtained, where the ferry-dependent coefficients are collected in the parameter $\check c_\mathrm p>0$. 
\begin{figure}[b!]
	\centering
	\includegraphics[width=0.47\textwidth]{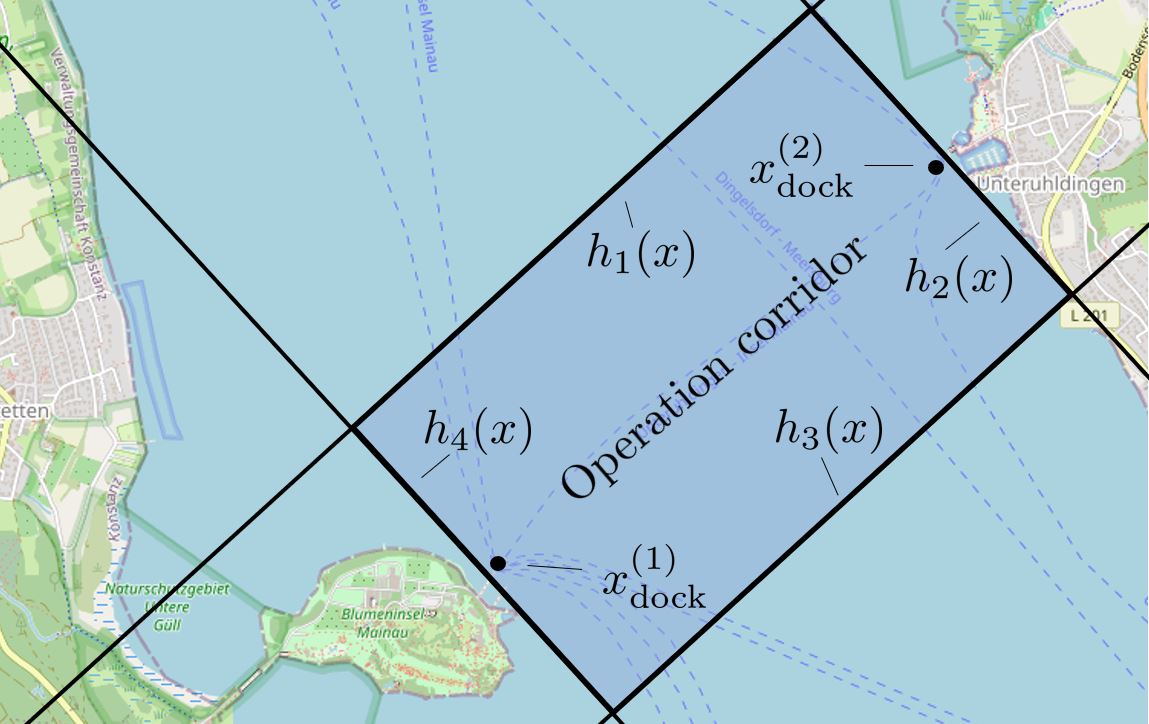}
	\caption{Map of the upper part of Lake Constance overlaid with the operation corridor including the docking positions. Map data ©2025 OpenStreetMap contributors, licensed under ODbL.}
	\label{fig_4} 
    
\end{figure}
\begin{figure}[t!]
	 \vspace*{0.2cm}
	\centering
	\includegraphics[width=0.5\textwidth]{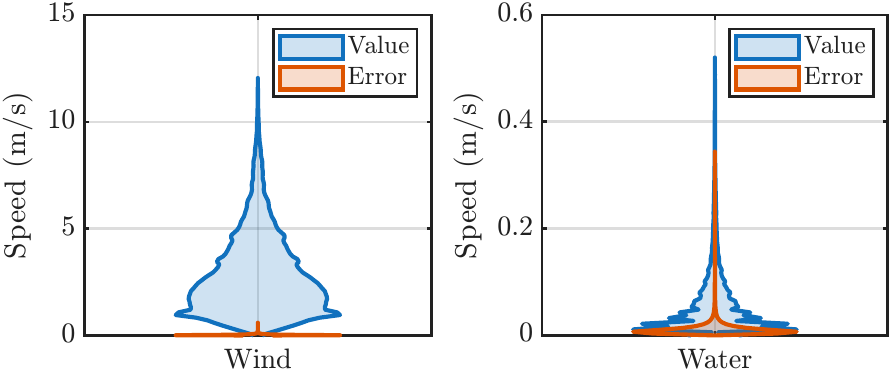}
	\caption{Violin plots of the PDE modeled current and wind speeds at the operation corridor and the errors of the simplified models.}
	\label{fig_5} 
    \vspace*{-0.4cm}
\end{figure}

\subsection{Path constraints}
The path constraints are used to limit the operation corridor and to keep the inputs within a reasonable range. 
To limit the operation area, we define a convex set of admissible 2D-positions by introducing $i=1,2,\ldots,I$ linear inequality constraints given by
$h_i(x)=s_ix_\mathrm{l}+q_iy_\mathrm{l}+c_i\leq0$
with parameters $s_i,q_i,c_i\in\mathbb{R}$.
Further, to limit the force of the azimuth thruster, the constraint
$ h_u(u)=X_\mathrm{a}^2+Y_\mathrm{a}^2-\overline F_\mathrm{AT}^2\leq 0 $ 
is chosen.
Based on the introduced functions, the path constraints are 
\begin{equation*}
    h(x,u)=[     h_1(x),\ldots,h_I{}(x),h_u(u)]^\top. 
\end{equation*}
The resulting operation corridor for the scenario considered in the next section is visualized in Fig.~\ref{fig_4}. Note that the constraints can be adapted easily to other scenarios.

\section{Identification and numerical results}\label{sec_exp}
 The different parts of the shrinking-horizon OCP are parameterized and identified in this section. Finally, a numerical solution approach and a simulation study are presented.
 
\subsection{Wind and current model}
 The spatial quadratic models of the current and wind conditions \eqref{eq_env_model} exhibit lower complexity compared to the tabular data representation \eqref{eq_tab_data}. To assess whether these models are adequate for capturing the environmental conditions, a descriptive data analysis of the historical wind and current conditions within the operation corridor, along with the corresponding model errors, is conducted. In total, over $2.1\times10^{6}$ data points are analyzed between November 3, 2024, and September 9, 2025. The results are summarized in the violin plots shown in Fig.~\ref{fig_5}. The maximum observed wind speed is $11.7\;\mathrm{m/s}$, corresponding to a Beaufort scale value of 6. The quadratic wind model yields a maximum deviation of $0.52\;\mathrm{m/s}$. The water current velocity does not exceed $0.51\;\mathrm{m/s}$, and detailed analysis reveals that 98.7\% of the water model data points exhibit errors below $0.05\;\mathrm{m/s}$. The wind model shows higher accuracy relatively because it is unaffected by coastal influences, unlike the water currents. The empirical results indicate that quadratic models are suitable choices in the given operating corridor. 
\begin{table}[b!]
\centering
\caption{Identified parameters of the \textit{MS Insel Mainau}.}
\begin{tabular}{c@{\hspace{0.5em}}r@{\hspace{0.5em}}l
                c@{\hspace{0.5em}}r@{\hspace{0.5em}}l
                c@{\hspace{0.5em}}r@{\hspace{0.5em}}l}
\hline
\multicolumn{3}{c}{Hydrodynamic}  & \multicolumn{3}{c}{Wind}     & \multicolumn{3}{c}{Other}  \\
\hline
$X_{u}$   & 1470   & Ns/m             & $A_\mathrm{Fw}$ & 59.2 & m$^2$                   & $m$                    & 35000 & kg  \\
$X_{uu}$  & 753    & Ns$^2$/m$^2$     & $A_\mathrm{Lw}$ & 219  & m$^2$                  & $\overline F_\mathrm{AT}$ & 24000 & N \\
$Y_{v}$   & 10290  & Ns/m             & $c_x$           & 0.59 & (-)                     & $\rho$                 & 1.204 & kg/m$^3$ \\
$Y_{vv}$  & 5272   & Ns$^2$/m$^2$     & $c_y$           & 0.84 & (-)                     & $\check c_\mathrm p$   & 0.0417 & m$^{1/2}$kg$^{-1/2}$ \\
\hline
\end{tabular}
\label{tab:parameters_ferry}
\end{table}
\subsection{Empirical parameter identification}
\begin{figure}[t!]
	 \vspace*{0.2cm}
	\centering
	\includegraphics[width=0.4\textwidth]{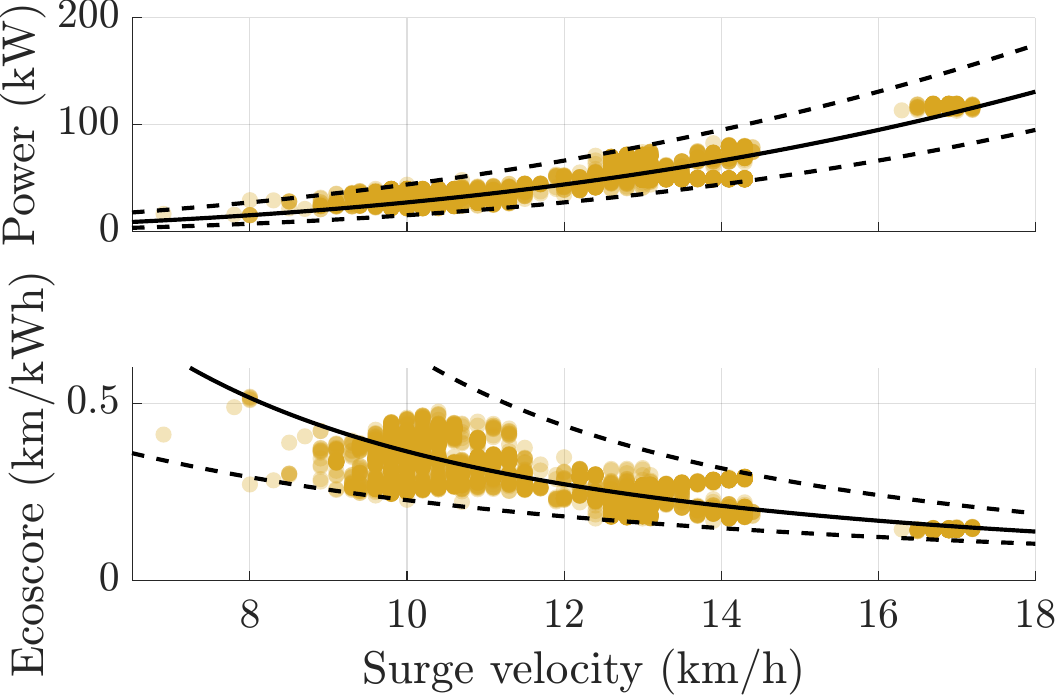}
	\caption{Power and \textit{ecoscore} of ferry \textit{MS Insel Mainau} over static surge velocity with parameters given in Table~\ref{tab:parameters_ferry} (black), dashed lines under uncertainties, and measured real-world data (gold).}
	\label{fig_6}  \vspace*{-0.4cm}
\end{figure}

The determined parameters of the dynamic model of the ferry \textit{MS Insel Mainau} are listed in Table~\ref{tab:parameters_ferry}. Linear and quadratic hydrodynamic damping coefficients are determined to match real-world measurements in least squares (cf. Fig.~\ref{fig_6}) in surge direction. The projected areas of the vessel are determined by its overall length of $30\;\mathrm{m}$, width of $8.1\;\mathrm{m}$, and height of $7.3\;\mathrm{m}$. The aerodynamic drag terms are determined by assuming an ellipsoidal body for the frontal effects and a cube with reduced height for the side effects \cite[Sec. 3]{Hoerner.1965}. Note that these parameters match the rough suggestion for vessel drag coefficients in \cite[Ch.~10]{Fossen.2021} and, if necessary, can be optimized by a detailed computational fluid dynamics (CFD) study. The power parameter $\check c_\mathrm{p}$ is fitted with least-squares to real-world data. Finally, the mass $m$, the maximum force $\overline F_\mathrm{AT}$, and the air density $\rho$ are prior known.   

\begin{figure}[b!]
	\vspace*{-0.2cm}
	\centering
	\subfigure{\includegraphics[height=5cm]{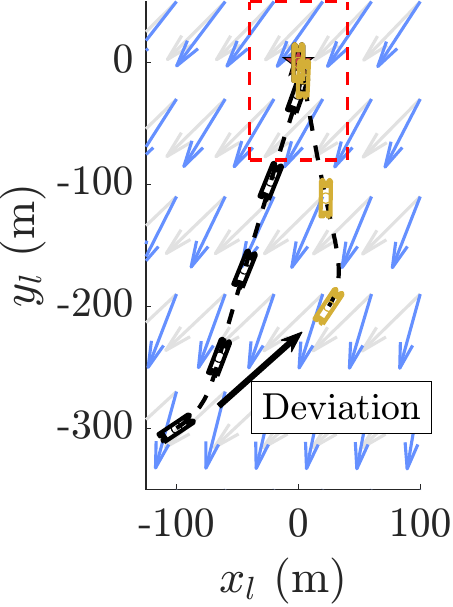}}
	\subfigure{\includegraphics[height=5cm]{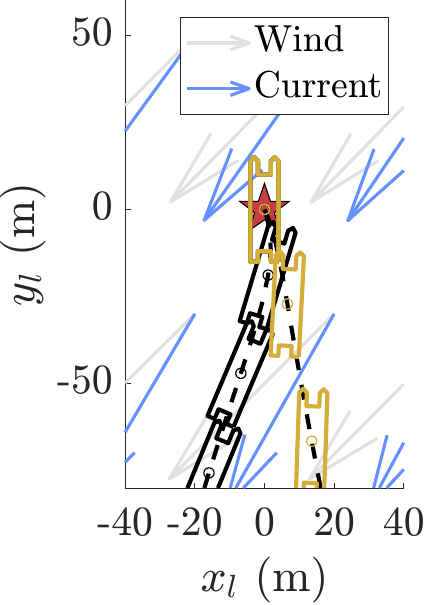}}
	\caption{Optimized ferry routes over the wind field and the current field with zoom on the right. The coordinate system is shifted and rotated to achieve $x_\mathrm{dock}^{(1)}=[0,0,\pi/2]^\top$ for visualization.}
	\label{fig_result}
\end{figure}

\subsection{Solution method and numerical results}
The OCP is transcribed in a nonlinear program (NLP) by direct multiple shooting \cite{Bock.1984}. The NLP is implemented via 
 {\fontfamily{qcr}\selectfont CasADi} \cite{Andersson.2019} and solved with {\fontfamily{qcr}\selectfont IPOPT} \cite{Wachter.2006}. For integration, the explicit Runge-Kutta method of 4$^\mathrm{th}$-order is employed. To illustrate the results, we consider two scenarios.
\paragraph*{Scenario 1} This scenario demonstrates the decision support in standard operation. For a wind speed of $11\;\mathrm{m/s}$ and spatially varying currents with amplitude $0.1\;\mathrm{m/s}$, the suggested trajectory with remaining
$T-t_1=240\;\mathrm{s}$ is shown in Fig.~\ref{fig_result} in black. Since the crew deviates from the initial plan, the trajectory is updated sequentially. The updated trajectory suggested for $T-t_2=120\;\mathrm{s}$ is plotted in gold. Both trajectories also include yaw angles and velocity profiles that support environment-aware and energy-efficient operation.
\paragraph*{Scenario 2}
A simulation study is conducted to examine the impact of environmental conditions and maneuver time on the energy required for the dock-to-dock transfer. A distance of $2527\;\mathrm{m}$ separates the docking positions (cf. Fig.~\ref{fig_4}). The Pareto fronts obtained with the proposed method are shown in Fig.~\ref{fig_8}.  Both the wind and the current are directed from the north, where 100$\%$ corresponds to a wind speed of $13\;\mathrm{m/s}$ and a current speed of $0.2\;\mathrm{m/s}$. The required energy increases with shorter dock-to-dock time and stronger environmental disturbances. This information can guide future time scheduling in ferry operations.
For reproducibility, the full implementation is available at: \url{https://github.com/hhomb/Ferry}
 
\begin{figure}[t!]
	 \vspace*{0.2cm}
	\centering
	\includegraphics[width=0.45\textwidth]{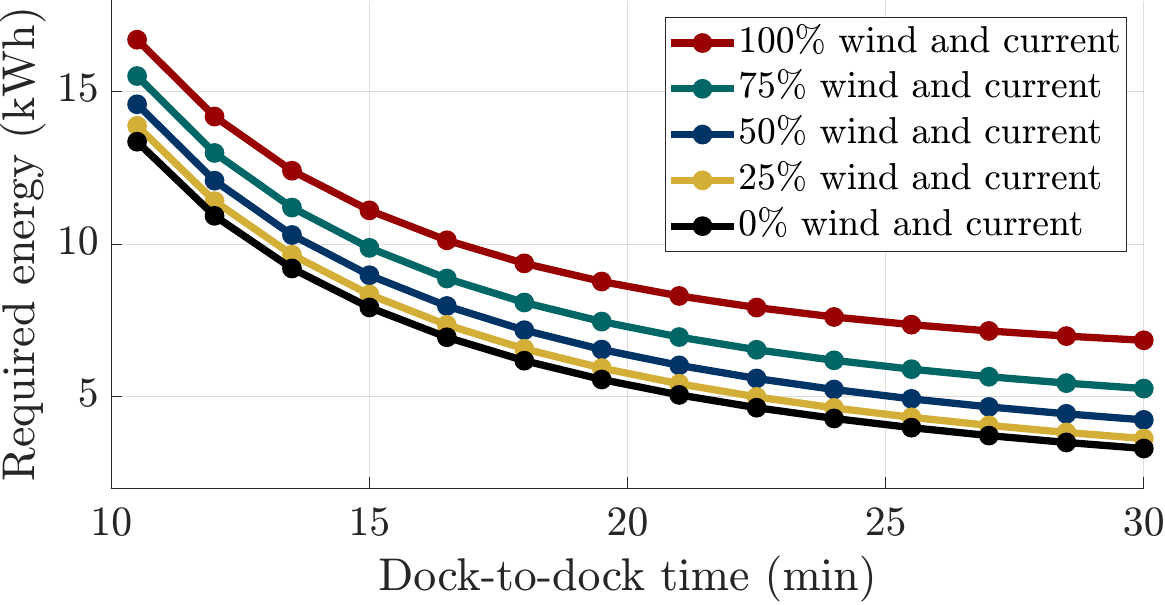}
	\caption{Pareto fronts given by required energy over maneuver time for different environmental conditions in \textit{Scenario 2}.}
	\label{fig_8}  \vspace*{-0.4cm}%
\end{figure}

\section{Conclusion}\label{sec_con}
This paper presents the design of a decision support system for the real-world ferry \textit{MS Insel Mainau} based on a dynamic model. 
The presented approach is formulated as a shrinking-horizon optimal control problem, demonstrating the potential of embedded numerical optimization as a powerful tool for developing semi-autonomous ferries that prioritize energy efficiency while fulfilling several additional requirements. 
The real-world data indicate a high potential to save energy through environment-aware trajectory optimization.
The resulting optimal control problem is solved numerically using a direct approach, and the solution trajectory meets the specifications.  
In future work, the decision support system can be attempted in the real world, and the results can be compared to the ground-truth with a complete {3-DOF} model.






\section*{Funding and acknowledgment}
This research was supported by the City of Konstanz' Smart Green City program as part of the "Model Projects Smart Cities" funding program of the German Federal Ministry for Housing, Urban Development and Building (BMWSB) under grant no. KfW 13622889 and  DFG via projects 504452366 (SPP 2364) and 525018088.
 The authors would like to thank Bodensee Schifffahrtsbetriebe (BSB), with Managing Director Christoph Witte and his team, for making the real-world ferry \textit{MS Insel Mainau} available for our research. Additionally, we appreciate the support of the Environmental Protection Agency of Baden-Württemberg (LUBW) for providing the estimated current and wind data.


%
%
%
%
%
%

\bibliographystyle{IEEEtran}
\bibliography{references}

\end{document}